\documentclass[preprint,aps,tightenlines,]{revtex4}
\usepackage{amsmath,amsfonts,amssymb,amsthm}
\usepackage{color}
\usepackage{graphicx,latexsym}

\begin{document}

\title{Coulomb-induced dynamic correlations in a double nanosystem}

\author{Valeriu Moldoveanu}
\affiliation{National Institute of Materials Physics, P.O. Box MG-7,
Bucharest-Magurele, Romania}
\author{Andrei Manolescu}
\affiliation{Reykjavik University, School of Science and Engineering, Menntavegi 1,
IS-101 Reykjavik, Iceland}
\author{Vidar Gudmundsson}
\affiliation{Science Institute, University of Iceland, Dunhaga 3, IS-107 Reykjavik,
Iceland}
\affiliation{Physics Division, National Center for Theoretical Sciences,
             PO Box 2-131, Hsinchu 30013, Taiwan}

\begin{abstract}

Time-dependent transport through two capacitively coupled quantum dots is
studied in the framework of the generalized master equation.  The Coulomb
interaction is included within the exact diagonalization method. Each
dot is connected to two leads at different times, such that a steady
state is established in one dot {\it before} the coupling of the other dot
to its leads.  By appropriately tuning the bias windows on each dot we
find that in the final {\it steady state} the transport may be suppressed
or enhanced.  These two cases are explained by the redistribution of
charge on the many-body states built on both dots.  We also predict and
analyze the {\it transient} mutual charge sensing of the dots.

\end{abstract}


\maketitle

\section{Introduction}

Recent on-chip measurements show how two nearby mesoscopic conductors
with little or no particle exchange interact via Coulomb forces.
For example a quantum point contact (QPC) has been used as a charge
detector (electrometer) near a quantum dot (QD) \cite{Field,Johnson}
and in measurements of the counting statistics of the electrons in
the dot \cite{Gustavsson}.  Conversely, the backaction of the current
flowing through the QPC on the states of the QD have been demonstrated
\cite{Onac,Sukhorukov}. A ratchet effect in a {\em serial} double quantum
dot (DQD) driven by the current in the nearby QPC has been recently
reported \cite{Khrapai}.  The electrons in the serial DQD could also be
excited by photons emitted by the QPC \cite{Gustavsson2,Ouyang} or by
phonons \cite{Gasser}.

Transport experiments in a {\em parallel} DQD with tunable coupling
have been performed by McClure {\em et al.}\ \cite{McClure}.  Both positive
and negative cross current correlations have been observed and related
to the interdot Coulomb interaction, whereas in the noninteracting
case only negative correlations are expected \cite{Buttiker}. Another
effect of the Coulomb correlations in parallel dots is the mesoscopic
Coulomb drag \cite{Goorden,EPL}.  Unlike the macroscopic drag effect
which is a result of quasi-equilibrium thermal fluctuations in the
drive circuit \cite{Levchenko}, the current in an unbiased dot is
driven by nonequilibrium time-dependent charge in the second, biased
dot \cite{Sanchez}.

In the theoretical descriptions of transport in parallel DQDs each
dot is coupled to two semi-infinite leads seen as particle reservoirs
with fixed chemical potentials.  When the lead-dot coupling is weak
(tunneling regime) rate or Markovian master equations are used
\cite{Ouyang,McClure,Sanchez,Welack,Stace}. Usually the dots are
considered one-level systems and the dot-dot interaction is reduced to
one parameter.  For strong lead-dot coupling a scattering theory has
been formulated \cite{Goorden} and also Keldysh-Green methods
combined with phenomenological interaction \cite{Levchenko} or with
the random-phase approximation \cite{EPL}.  Most of the theoretical
calculations were performed for the steady state.

In this paper we theoretically investigate Coulomb correlation
effects in capacitively coupled parallel nanosystems {\it both} in
transient and steady states regime. Conventionally we shall call them
quantum dots, but the method we use is adaptable to any sample geometry
and any number of leads.  In our setup each QD is connected to the leads
at different moments and due to the Coulomb interaction they mutually
respond to each other's transient charging or discharging. The aim of
this work is to describe and understand these effects.  Depending on the
initial conditions (occupations, bias voltages) the current cross-correlations
may be positive or negative.  The calculations are performed within the
generalized master equation (GME) method for the reduced density operator
(RDO)  of the double dot.  The formalism was adapted for open mesoscopic
systems by several authors \cite{GMEmeso}.  We used it recently to study the
transient  behavior of open noninteracting \cite{NJP1,NJP2,PRB1} and
interacting nanosystems \cite{PRBCB}.  The interaction is treated
with the exact diagonalization method, both intradot and interdot, on
equal footing.

\section{Theory}

\begin{figure}[tbhp!]
\includegraphics[width=0.40\textwidth]{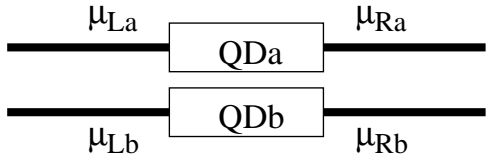}
\caption{The system: A double dot and four leads.}
\label{system}
\end{figure}

The Hamiltonian of the total system, shown in Fig.\ \ref{system}, is 
\begin{equation}
H(t)=H_{\rm S}+H_{\{l\}}+H_{\rm T}(t), 
\end{equation}
where S stands for the ``sample'', 
in this case the DQD, i.e. ${\rm QD}_a+{\rm QD}_b$, and 
$\{l\}= \mathrm{\{L_a,R_a,L_b,R_b \}}$ is the set of leads. 
$H_{\rm T}$ incorporates the sample-leads tunneling,
\begin{equation}\label{Htunnel}
H_{\rm T}(t)=
\sum_{n}\sum_{l}\int dq \chi_{l}(t)
\left( T^l_{qn}c^{\dagger}_{ql}d_n+h.c.\right) ,
\end{equation}
with $\chi_{l}(t)$ time-dependent functions describing the 
contact with the lead $l$.  $c^{\dagger}_{ql}/d_n$ are the
creation/annihilation operators in the leads and sample respectively, 
and $T^{l}_{qn}$ are model specific coupling coefficients..
%
The RDO $\rho(t)$, or the ``effective'' statistical operator of the
open sample, is defined by averaging the statistical operator of the
total system over the states of all leads.  In the lowest (quadratic)
order in $H_T$ it satisfies the GME.
\begin{eqnarray}
\label{GMEfin}
&&{\dot\rho}(t)=-\frac{i}{\hbar}[H_{\rm S},\rho(t)]  \\
&&-\frac{1}{\hbar^2}{\rm Tr}_{\{l\}}
\left[ H_{\rm T}(t),\int_{t_0}^t ds U_{t-s} \left[H_{\rm T}(s),\rho(s)
\rho_{\{l\}}\right] U_{t-s}^{\dagger}  \right] ,  \nonumber \\
\nonumber
\end{eqnarray}
where $U_{t}=e^{-it(H_S+H_{\{l\}})/\hbar}$ is the evolution operator of
the disconnected system, and $\rho_{\{l\}}$ is the statistical operator of
the leads in equilibrium which is the product of the Fermi distributions
of each lead $l$ with chemical potential $\mu_l$.  Before the leads are
coupled $\rho(t)$ describes an equilibrium state of the isolated sample
\cite{NJP1,NJP2,PRB1,PRBCB}.

$H_S$ also includes the Coulomb interaction. The interacting many-electron
states (MES) of the isolated sample, solutions of $H_S | \alpha \rangle =
\cal E_{\alpha} | \alpha \rangle$, are found by exact diagonalization.
Each MES is expanded in the Fock space built on a finite number of
single-electron states (SES), $N_{SES}$.  The number of electrons
$N$ in the sample may vary between zero and $N_{SES}$ and hence the
number of MES is $N_{MES}=2^{N_{SES}}$.  Since the dots are not in
direct tunneling contact the number of electrons in each dot, $N_a$
and $N_b$, respectively, are ``good quantum numbers'' for the MESs.
The ground-state energies of the isolated DQD can be labeled as ${\cal
E}_g(N_a,N_b)$.  The chemical potential of a MES with $N=N_a+N_b$
electrons, $\mu(N_a,N_b)$, is the energy cost to add one more electron
to the ground-state with $N-1$, and has to fit with the leads' chemical
potentials in order to allow transfer of electrons.

We solve Eq.\ (\ref{GMEfin}) numerically in the MES basis $\{ | \alpha
\rangle \}$.  Using the RDO we can calculate the mean number of electrons
and hence the charge in each dot, $Q_i(t),\ (i=a,b)$, and by taking the
time derivative we obtain the currents in each lead,
\begin{equation}
\label{occup}
\dot Q_i(t)=\sum_{n_i} n_i\sum_{\alpha_{n_i}}\dot\rho_{\alpha_{n_i}\alpha_{n_i}} 
=J_{{\rm L}i}(t)-J_{{\rm R}i}(t),
\end{equation}
where $\alpha_{n_i}$ are the MESs of the double system with $n_i$
electrons in ${\rm QD}_i$.  We can thus describe the partial charge
and currents associated with any partition of electrons.  The currents
corresponding to each lead are identified from the last term of Eq.\
(\ref{GMEfin}) \cite{NJP1,PRBCB}.  A current is positive when flowing
from left to right and negative otherwise.

\section{Results}

We use a lattice model for our system, each QD being a chain of four
sites.  The electrons are distributed on a $2\times 4$ lattice, but
the hopping between the chains is forbidden.  
The coupling coefficients are 
\begin{equation}
T^{l}_{qn}=V_0{\psi}^{*}_{ql}(0)\phi_n(i_l)\,
\end{equation}
where $\psi_{ql}$
and $\phi_n$ being the single-particle wave functions in the leads and 
in the sample, evaluated at the contact sites labeled as 0 and $i_l$, 
respectively \cite{NJP1}.  The parameter $V_0$ gives the coupling strength.
All electrons on this lattice
interact with pairwise Coulomb potentials $U/d_{jk}$ with $d_{jk}$ the
distance between electrons $j$ and $k$ and $U$ a strength parameter.
Coulomb forces are neglected in the leads.  We use all 8 SES of the lattice
to calculate all 256 MES, and the first 40 MES are sufficient to obtain convergent
results from Eq.\ (\ref{GMEfin}) \cite{PRBCB}.  Our energy unit is the
hopping energy in the dots $t_D$, the time unit is $\hbar/t_D$, and the
currents are calculated in units of $et_D/\hbar$.  We use $V_0=1.5$ 
and $U=1$.
The ground-state energies for the DQD are ${\cal E}_g(0,0)=0,\
{\cal E}_g(1,0)=2.38,\ {\cal E}_g(1,1)=5.47,\ {\cal E}_g(2,0)=6.36,
\ {\cal E}_g(2,1)=10.08,\ {\cal E}_g(3,0)=12.37, \ etc.$, 
and thus $\mu(1,1)=3.09,\ \mu(2,0)=3.97,\ \mu(2,1)=4.6$.  
The dots being identical ${\cal E}_g(N_a,N_b)={\cal E}_g(N_b,N_a)$.

\begin{figure}[tbhp!]
\centering
\includegraphics[width=0.74\textwidth]{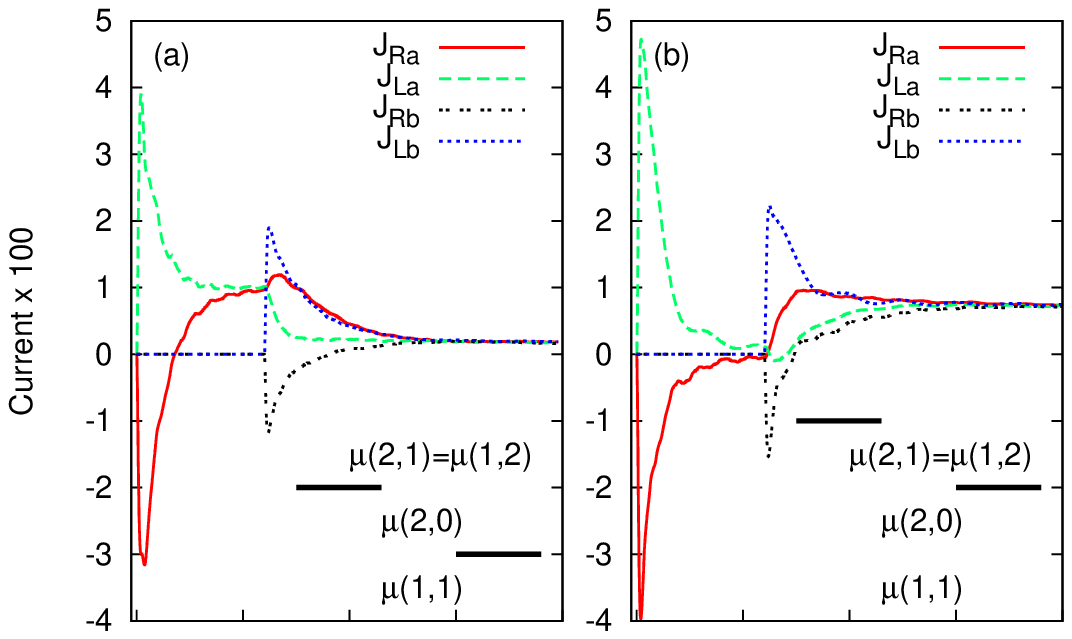}
\vskip -3cm
\includegraphics[width=0.74\textwidth]{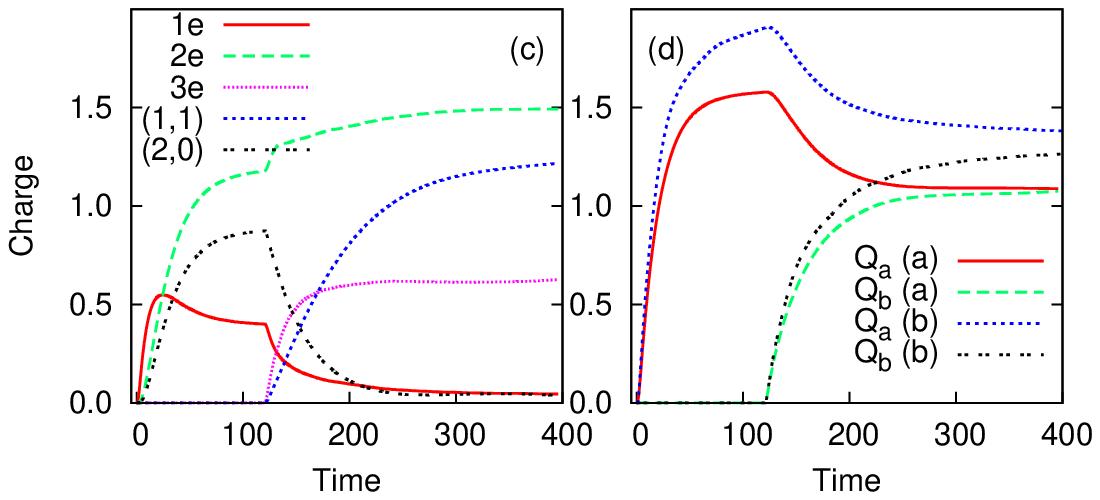}
\vskip -2.0cm
\caption{(Color online) (a)-(b): The total currents in the leads
for two bias windows: 
(a) $\mu_{\rm La}=\mu_{\rm Lb}=4.25$, $\mu_{\rm Ra}=\mu_{\rm Rb}=3.75$; 
(b) $\mu_{\rm La}=\mu_{\rm Lb}=4.75$, $\mu_{\rm Ra}=\mu_{\rm Rb}=4.35$.
The insets show the bias windows and the MES chemical potentials. 
(c) Partial charge for the states involved in (a) (see text);
(d) Total charge on each dot for (a) and (b).
}
\label{totcrt}
\end{figure}

In the following cases both dots are initially empty and $\mu_{\rm
La}=\mu_{\rm Lb}, \mu_{\rm Ra}=\mu_{\rm Rb}$.  $\mathrm{QD_a}$ opens at
$t_a=0$ and after a charging period it evolves towards a steady state.
In Fig.\ \ref{totcrt}(a,b) we show the current in $\mathrm{QD_a}$
for two choices of the chemical potentials of the leads.  In the first
case $\mu(1,1)<\mu_{\rm Ra}<\mu(2,0)<\mu_{\rm La}<\mu(2,1)$, meaning
that in the steady state of $\mathrm{QD_a}$ the main contributor to the
current is the MES $(2,0)$ \cite{PRBCB}.  $\mathrm{QD_b}$ is coupled at
$t_b=120$ when a new transient period begins for both dots, after which
all currents end up at equal values, considerably smaller than before
$t_b$.  So one can say the two dots are {\em negatively} correlated:
The activation of one inhibits the other until they block each other,
Fig.\ \ref{totcrt}(a).  In the second case, Fig.\ \ref{totcrt}(b),
we have instead $\mu(2,0)<\mu_{\rm Ra}<\mu(2,1)<\mu_{\rm La}$ and
only a very small current passes through $\mathrm{QD_a}$ in the first steady
state due to the Coulomb blockade. But the coupling of $\mathrm{QD_b}$
now activates $\mathrm{QD_a}$, so the dots become {\em positively}
correlated \cite{McClure}.

To explain what is going on we show in Fig.\ \ref{totcrt}(c) the
population of the relevant states for the first case, calculated with
Eq.\ (\ref{occup}).  As long as $\mathrm{QD_b}$ is closed one- and
two-particle states of $\mathrm{QD_a}$ are charging yielding a total
charge up to $Q_a\approx 1.5$, as further shown in Fig.\ \ref{totcrt}(d).
Once $\mathrm{QD_b}$ opens the electrons tunneling into it repel
some charge from $\mathrm{QD_a}$ and new MESs are being created like
(1,1) and (2,1).  Since ${\cal E}_g(2,0)>{\cal E}_g(1,1)$ the new
two-particle ground state is (1,1) and hence the transition $(2,0)\to
(1,1)$ occurs, but also $(2,0)\to (2,1)$.  The later is possible because
${\cal E}_g(2,1)-{\cal E}_g(2,0)=3.72$ is slightly below the bias window.
Consequently the states (2,0) depopulate fast whereas the populations of
the states (1,1) and $(2,1)$ (and $(1,2)$ as well) increase, as seen in
Fig.\ \ref{totcrt}(c). In the steady state the bias window is nearly empty
of any MES chemical potential and consequently the currents nearly vanish.
This is an interdot Coulomb blocking effect \cite{McClure}.  The total
charge in the dots converges to 2.2 electrons. Of that 1.5 reside on
two-electron states: 1.2 on the ground state (1,1), i.e.\  below the bias
window, and 0.3 on excited states (1,1) and MESs (2,0).  Also, about 0.6
electrons are on three-particle states (2,1), i.e.\  above the bias window.

\begin{figure}[htbq]
\centering
\includegraphics[width=0.74\textwidth]{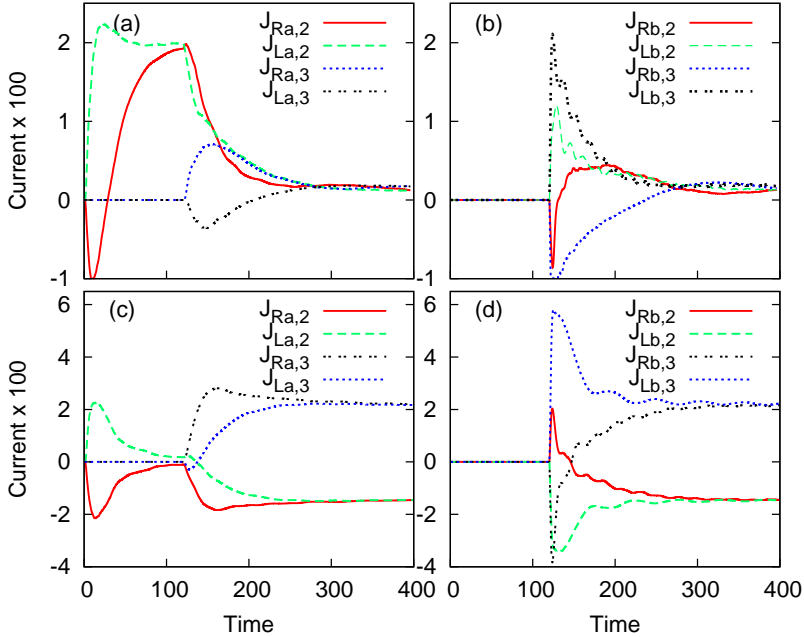}
\caption{(Color online) The partial currents carried by the 
two- and three-particle states in $\mathrm{QD_a}$ and $\mathrm{QD_b}$: 
(a-b) corresponding to Fig.\ \ref{totcrt}(a); 
(c-d) corresponding to Fig.\ \ref{totcrt}(b).
}
\label{parcrt1}
\end{figure}

In Fig.\ \ref{parcrt1}(a) we show the partial currents in the leads
connected to $\mathrm{QD_a}$ carried by the two- and three-particle states.
The former drop fast after $t_b$ during the depletion of the MES (2,0). 
Because $\mu(2,0)$ is almost in the center of the bias window   
$J_{\rm Ra,2}$ and $J_{\rm La,2}$ are very similar.  
The three-particle currents in $\mathrm{QD_a}$ are more interesting. They
correspond to the states (2,1) and, surprisingly, $J_{\rm La,3}<0$ and
$J_{Ra,3}>0$, meaning that $\mathrm{QD_a}$ ejects charge in both leads ${\rm
L}_a$ and ${\rm R}_a$.  The resulting ``lobe'' shape is also seen in Fig.\
\ref{totcrt}(a).  The net charging of the (2,1) states is actually done
through the leads connected to $\mathrm{QD_b}$, as can be seen in  Fig.\
\ref{parcrt1}(b), where $J_{\rm Lb,3}>0$ and $J_{\rm Rb,3}<0$, i.e.
both currents flow into the dot.

We return now to the positive correlation case.  For $t<t_b$ no chemical
potential of type $\mu(N_a,0)$ is inside the bias window and hence
no current flows in the steady state of $\mathrm{QD_a}$. The charging
goes up to $Q_a\approx 1.9$ as seen in Fig.\ \ref{totcrt}(d), with the
ground state (2,0) occupied.  When $\mathrm{QD_b}$ is open the new states
(2,1) are created 
and since the corresponding $\mu(2,1)$ is inside the bias window they are
available for transport.  After the transient phase, when $\mathrm{QD_b}$
is charging and $\mathrm{QD_a}$ is discharging, all currents reach
the same steady value, driven by the states (2,1) and (1,2) which end
up equally populated. The currents in the steady state have now two-
and three-particle components. These partial currents, shown in Fig.\
\ref{parcrt1}(c), have another curious behavior. Both $J_{\rm La,3}$ and
$J_{\rm Ra,3}$ are positive in the steady state, whereas $J_{\rm La,2}$
and $J_{\rm Ra,2}$ are negative, the net result being the total, positive
current.  This means that three-particle currents flow from left to right,
but the two-particle currents go from right to left. The reason is that
in our model the electrons are created or annihilated one at a time.
A MES (2,1) is formed by creating one more electron to the ground MES
(1,1), and so the positive (2,1) and (1,2) currents
deplete the (1,1) states.  But $\mu(1,1)$ is below the bias window and
so the (1,1) states have to be backfed by a negative, two-particle
current. The
single particle states are not occupied and do not contribute to transport.
The current in the circuit $a$ is carried by MES (2,1), but
not by (1,2), and the other way round in the circuit $b$.  The electrons
tunneling from the left leads $a$ and $b$ thus compete each other to
access the (1,1) MES. In turn, when electrons leave a dot the remaining
two-particle MES has the lowest energy ${\cal E}_g(1,1)$ and not ${\cal
E}_g(2,0)$ which is higher.  Such transitions are called ``U-sensitive
processes'' in Ref.\ \onlinecite{McClure}.

\begin{figure}[htbq]
\vskip -0.5cm
\centering
\includegraphics[width=0.74\textwidth]{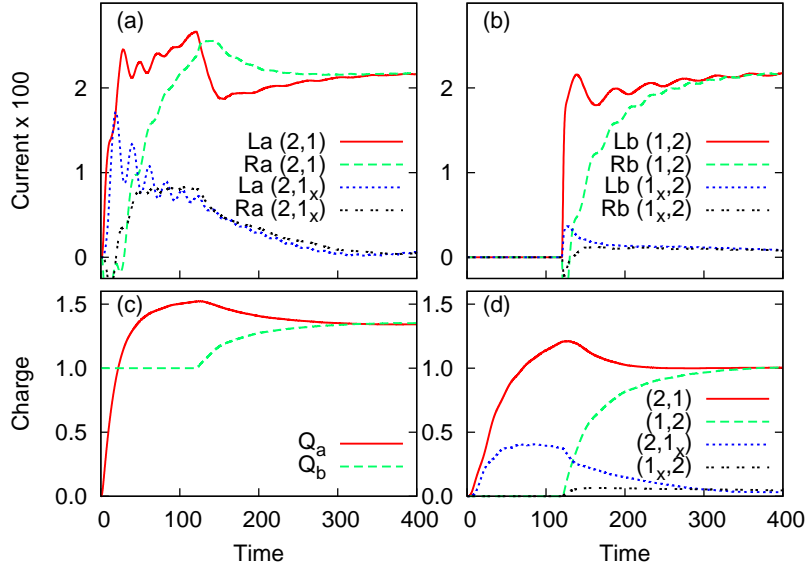}
\caption{(Color online) 
(a-b) The contributions of the ground state (2,1) and of the excited states 
$(2,1_x)$ to the currents in the leads.  $\mu_{\rm La}=\mu_{\rm Lb}=4.75$, 
$\mu_{\rm Ra}=\mu_{\rm Rb}=4.35$. $\mathrm{QD_b}$ initially contains one 
electron.
(c) The evolution of total charge in each dot. 
(d) Populations of the three-particle states. 
}
\label{parcrt2}
\end{figure}

The partial currents of states (2,1) and (1,2) are shown in Fig.\
\ref{parcrt2}(a,b).  In this case we use the same chemical potentials
as in Fig.\ \ref{totcrt}(b), but now $\mathrm{QD_b}$ contains one electron
in the ground state at $t=0$.  After $t=t_b$ $\mathrm{QD_b}$ absorbs more
charge and the double system evolves toward the same steady state as
before, Fig.\ \ref{parcrt2}(c). But prior to $t_b$, although isolated, 
the initial electron is being excited by the charging of $\mathrm{QD_a}$.
This can be seen in the Fig.\ \ref{parcrt2}(d) where the populations
of the ground state (2,1) and of the MESs containing the excited state
of the electron in $\mathrm{QD_b}$  denoted as $(2,1_x)$ are displayed.  
The currents in the circuit $a$ feel the initial electron in $\mathrm{QD_b}$, 
but also the excited states of it.  Indeed the MESs $(2,1_x)$ decay 
while the system approaches the steady state.
\begin{figure}[htbq]
\vskip -1.5cm
\centering
\includegraphics[width=0.74\textwidth]{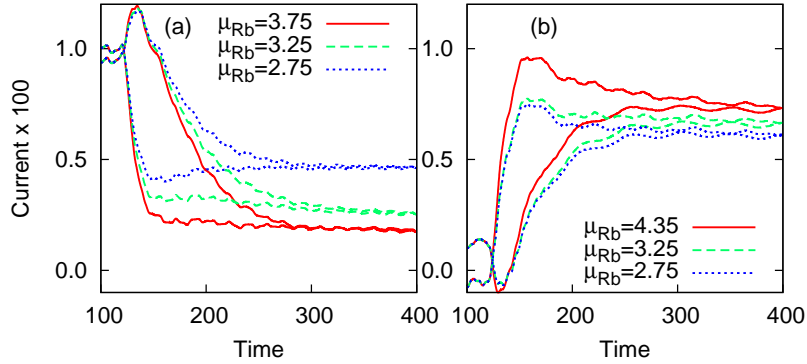}
\vskip -2cm
\caption{(Color online) 
The effect of the bias applied on $\mathrm{QD_b}$ on the currents $J_{\rm La,Ra}$. 
(a) $\mu_{\rm La}=\mu_{\rm Lb}=4.25, \mu_{\rm Ra}=3.75$.
(b) $\mu_{\rm La}=\mu_{\rm Lb}=4.75, \mu_{\rm Ra}=4.35$.
The same line type is used for the left and right currents.
}
\label{Imu} 
\end{figure}

Next we keep $\mu_{\rm Lb}$ fixed and decrease $\mu_{\rm Rb}$
relatively to the setup of Fig.\ \ref{totcrt}(a), for increasing the
bias $eV_b=\mu_{\rm Lb}-\mu_{\rm Rb}$. Fig.\ \ref{Imu}(a) shows that
the splitting of the two currents in $\mathrm{QD_a}$ during the second
transient phase decreases.  The final current increases with $V_b$,
but it is still smaller than the steady value before $t_b$. The
effect of increasing $V_b$ on the final currents occurs in two steps.
First the states (2,0) and (0,2) become slightly populated and tunneling
of one electron creates three-particle currents, Fig.\ \ref{Imu}(a)
with $\mu_{\rm Rb}=3,25$.  Then, the bias window approaches $\mu(1,1)$
and eventually includes it, and tunneling on MES (1,1) amplifies
the three-particle currents.  Since $V_b$ is acting directly on
$\mathrm{QD_b}$ the final currents in the $b$ circuit are larger than in
$a$ (not shown).  Fig.\ \ref{Imu}(b) shows the result of increasing $V_b$
starting with the setup of Fig.\ \ref{totcrt}(b).  The Coulomb blockade
on $\mathrm{QD_a}$ is still lifted when the bias on $\mathrm{QD_b}$
increases.  The discharging in the $a$ arm may be large enough to produce
a {\em negative} left current.

Finally, one comment on the interdot distance. The interdot Coulomb
interaction decreases with the distance between the dots, but for
simplicity we kept it equal to the lattice constant.  Increasing it the
MESs change, but similar effects were obtained by appropriately tuning
the chemical potentials of the leads.

\section{Conclusions}

In conclusion, we discussed time dependent charge sensing effects
and computed mutually sensitive currents in parallel quantum dots.
A steady-state transport regime of one dot is suppressed after connecting
the second one. Conversely, the current through one dot increases if
the charging of the second dot opens new many-body channels within the
bias window.  In particular, we predict that the transient current in
the leads attached to the first dot may change sign when the second dot
is connected.  This effect can be experimentally tested.

The RDO of the coupled system and the GME describe its entangled dynamics
by treating all electrons equally.  The Coulomb effects are fully
included and the charging and discharging energies are present in the
MES structure. The classical charging and the quantum correlations are
treated together. The exact many-body states fit naturally with the 
Fock space formulation of the GME. The access to individual MES allows
a better understanding of the Coulomb-induced effects on the total
 currents. 
 
\begin{acknowledgments} This work was supported by the Development Fund
of Reykjavik University (grant T09001), the Icelandic and the University of Iceland 
Research Funds, and the Romanian Ministry of Education and Research 
(grants PNCDI2 515/2009 and 45N/2009).  
\end{acknowledgments}

\bibliographystyle{apsrev}

\end{document}